\documentclass[pra,twocolumn]{revtex4}

\usepackage[dvips]{graphicx}
\usepackage{ifthen}
\usepackage{amstext}
\usepackage{amsmath}
\usepackage{amsfonts}
\usepackage{amssymb}

\newcommand{\ket}[1]{|#1\rangle}
\newcommand{\bra}[1]{\langle#1|}
\begin{document}
\title{Eavesdropping on practical quantum cryptography}
\author{Mark Williamson and Vlatko Vedral}
\affiliation{Optics Section, Blackett Laboratory, Imperial
College, London SW7 2BW, UK}
\date{\today}
\begin{abstract}
Practical implementations of quantum cryptography use attenuated
laser pulses as the signal source rather than single photons. The
channels used to transmit are also lossy. Here we give a simple
derivation of two beam-splitting attacks on quantum cryptographic
systems using laser pulses, either coherent or mixed states with
any mean photon number. We also give a simple derivation of a
photon-number splitting attack, the most advanced, both in terms
of performance and technology required. We find bounds on the
maximum disturbance for a given mean photon number and observed
channel transmission efficiency for which a secret key can be
distilled. We start by reviewing two incoherent attacks that can
be used on single photon quantum cryptographic systems. These
results are then adapted to systems that use laser pulses and
lossy channels.

\end{abstract}
\maketitle
\section{Introduction}

Quantum cryptography is, in theory, unbreakable \cite{GisinRev,
SciAm}. The laws of quantum mechanics allow a sender of a message
and a receiver, known as Alice and Bob respectively, to defeat an
eavesdropper, Eve, from accessing their communications. In the
first quantum cryptography protocol discovered in 1984 by Bennett
and Brassard (BB84) \cite{BB84} the idea is that Alice encodes the
logical bit values $0$ and $1$ on to four possible polarisation
states of a photon, two in one basis such as horizontal and
vertical polarisation ($\ket{x}$ and $\ket{y}$ respectively) and
two in the conjugate basis, an equal superposition of the first
basis. These photons could be encoded as having $45^o$ and $-45^o$
polarisation ($\ket{u}$ and $\ket{v}$ respectively). Alice
randomly sends one of them to Bob who will then measure each of
the photons in one of the two bases also randomly and
independently from Alice. After Bob has made his measurement,
Alice reveals on a public channel the basis in which she sent her
photon. Bob then tells her whether or not he measured in the same
basis. If he did not or he did not detect a photon, the bit is
discarded but if the same basis was used they keep the bit and
both parties should now hold the same result of the bit's value.
The procedure is repeated until Alice and Bob hold a long string
of bits. This is their sifted key and they should have perfectly
correlated bit strings. If, however, a malicious eavesdropper,
Eve, was listening in an effort to obtain some of the key's
information, she must make measurements on the states Alice sends
to Bob. In doing this she will sometimes choose the wrong
measurement and create errors in Alice and Bob's sifted key
providing a signature of her presence to Alice and Bob when they
come to a process called error correction and estimation. Alice
and Bob sacrifice a subset of their key in order to do this and
from their estimate of the error rate they may also calculate the
maximum amount of information an eavesdropper could have obtained,
a limit bounded by the laws of physics but as yet still unknown.

Bounds for specific attacks have been found; these are usually
attacks in which the eavesdropper measures each individual state
Alice sends to Bob known as incoherent attacks. Incoherent attacks
on the BB84 protocol usually have the added restriction that they
take into account information gained by the eavesdropper after
public announcement of the basis but before error correction and
privacy amplification. The optimal bound of an unrestricted attack
using collective measurements of any size is still unknown. These
may not be severe limits to the security of quantum cryptography
at present because the technology needed to perform the simplest
of controlled interactions on more than a few qubits is still at
an early stage.

In reality however, there are added complications. All errors do
not generally come from an eavesdropper alone. There will be noise
in Bob's detectors and in the quantum channel they use to setup
the key as well as equipment misalignments. Real implementations
of quantum cryptography also have losses during transmission and
they do not at present use single photon sources. Instead,
attenuated laser pulses that may contain $0$, $1$, $2$ or more
photons are used as Alice's signal source. Practical
implementation of quantum cryptography opens a security loophole
and Eve may adapt her attacks to take advantage of these facts. We
give a simple derivation of incoherent eavesdropping attacks on
these imperfect systems. We give a general argument that applies
equally well to systems with high mean photon number pulses as we
make no approximation. These results apply equally well to
coherent states as to mixed photon number states with Poissonian
distribution. Three attacks are analyzed in increasing order of
technological sophistication of the eavesdropper. This is another
important point: the practical quantum cryptography system need
only prevent the most technologically powerful eavesdropper at the
time key distribution takes place. This is unlike classical
cryptography where a message may be stored until an eavesdropper
has the technology with which to break it \cite{GisinRev}.

Provided Alice and Bob can calculate the maximum level of
information Eve could have obtained on their key from error
estimation and by making a decision on how advanced her attack
could have been in the worst case, Alice and Bob may be still able
to distill a secret key on which Eve has negligible information
using a classical process called privacy amplification
\cite{PrivAmp}. Finding these limits is important for the security
of quantum cryptography which \emph{could} take into account the
technology of the eavesdropper and \emph{should} include the
settings and limits of Alice and Bob's equipment as well as
fundamental physics. We give bounds on the maximum error rate
Alice and Bob can tolerate for a realistic system with channel
losses and laser pulses as the signal source for three incoherent
attacks increasing in the technological power of the eavesdropper.

The paper is organized as follows. We start by reviewing
eavesdropping attacks where Alice uses a single photon source to
send key bits to Bob in section~\ref{ais}. We review two
strategies: The simplest strategy that Eve can employ, that of
intercept-resend, a strategy whose implementation is well within
the reach of current technology to that of the optimum incoherent
attack for which the technology is not yet available. This latter
attack involves storing quantum states for long times and well
controlled interactions between a probe (one or more qubits) and
the state the eavesdropper is trying to identify. These strategies
provide the foundations for section~\ref{ars}, eavesdropping on
systems that use laser pulses and have a quantum channel with a
transmission efficiency that may be less than unity. Beam-splitter
attacks are analyzed in detail using intercept-resend or optimal
incoherent attacks. Lastly, the photon-number splitting attack is
analyzed, the most advanced strategy an eavesdropper could use and
relies on performing unlimited efficient quantum non-demolition
measurements: technology that is unlikely to be available for a
long time. In sections~\ref{countermeasures} and
\ref{countermeasures2} we present some analysis on what Alice and
Bob can do to counter Eve's attacks.

\section{Incoherent eavesdropping: single photon sources}\label{ais}

Presented here are two attacks that an eavesdropper, Eve, can
employ depending on the sophistication of her technology on a
perfect quantum cryptographic system using the BB84 protocol. The
simplest, an intercept-resend strategy, could be implemented with
equipment currently available whereas an attack using the optimal
incoherent strategy would require an advance in technology. The
latter attack belongs to class sometimes referred to as
`translucent attacks with entanglement' \cite{EkertEve}.

\subsection{The intercept-resend strategy}\label{idealI-R}

The idea with this kind of attack is to measure all or a
proportion, $\epsilon$, of the states Alice sends to Bob. If she
chooses to measure only a fraction, $\epsilon$, of the states, how
she chooses which states to measure is dependent on how much
information she wishes to obtain on the final message while making
her presence as inconspicuous as possible. We will not consider
how she chooses this fraction.

Presented here is an argument adapted from \cite{Huttner,
Bechmann-Pasquinucci}. Eve makes her measurement on a state and
given her result she makes the best guess as to what the
\emph{logical bit value} of the state was that Alice intended Bob
to receive. Eve then prepares one of the two states in the
intermediate or Breidbart basis depending on this guess and sends
it on to Bob. As we shall see, the two intermediate states belong
to the same basis in which Eve makes her measurement.

The four states Alice can send to Bob in the BB84 protocol are
\begin{displaymath}
\ket{x}=\frac{1}{\sqrt{2}}(\ket{u}+\ket{v})
\end{displaymath}
\begin{displaymath}
\ket{y}=\frac{1}{\sqrt{2}}(\ket{u}-\ket{v})
\end{displaymath}
\begin{displaymath}
\ket{u}=\frac{1}{\sqrt{2}}(\ket{x}+\ket{y})
\end{displaymath}
\begin{equation}
\ket{v}=\frac{1}{\sqrt{2}}(\ket{x}-\ket{y})
\end{equation}
Eve does not have to guess the state with the best probability of
being correct. Alice and Bob are trying to set up a key composed
of 1s and 0s. It does not matter what state it is if Eve is only
trying to find out the values of the bits. Alice must encode her
states as in figure~\ref{logicalBB84}. Note that the logical bit
values for the two states in the same basis must be different
otherwise public announcement of the basis would result in Eve
knowing everything about the key by simply passively monitoring
this public conversation. Eve's best chance of guessing the value
of the bit with a simple, presently implementable measurement is
to have two outcomes $M_0$ and $M_1$. The probability of guessing
the bit value correctly is
\begin{equation}\label{Pcgeneral}
P_c=\frac{1}{4}[\langle x|M_0|x\rangle + \langle v|M_0|v\rangle +
\langle y|M_1|y\rangle + \langle u|M_1|u\rangle]
\end{equation}
Where the two measurement operators have to satisfy $M_0+M_1=I$.
$I$ the identity matrix. The measurement operators are
\begin{displaymath}
M_0=\ket{0}\bra{0}
\end{displaymath}
\begin{equation}
M_1=\ket{1}\bra{1}
\end{equation}
The two states $\ket{0}$ and $\ket{1}$ are given by
\begin{equation}\label{naughtandone}
\ket{0}=\cos\theta\ket{x}-\sin\theta\ket{y}
\end{equation}
\begin{equation}
\ket{1}=\sin\theta\ket{x}+\cos\theta\ket{y}
\end{equation}
Where $\theta$ is a variable angle taken clockwise from $\ket{y}$.
Substituting the forms of $M_0$ and $M_1$ into
equation~(\ref{Pcgeneral}) we get
\begin{figure}
\centering
\begin{picture}(200,220)
\put(50,100){\vector(0,1){100}}

\put(50,100){\vector(1,0){100}}

\put(50,100){\vector(1,1){71}}

\put(50,100){\vector(1,-1){71}}

\put(46,205){{\tt 1}}

\put(46,218){$\ket{y}$}

\put(155,97){{\tt 0}}

\put(165,97){$\ket{x}$}

\put(125,170){{\tt 1}}

\put(130,180){$\ket{u}$}

\put(125,25){{\tt 0}}

\put(130,15){$\ket{v}$}
\end{picture}
\caption{The four states of the BB84 protocol represented by their
logical bit values.}\label{logicalBB84}
\end{figure}
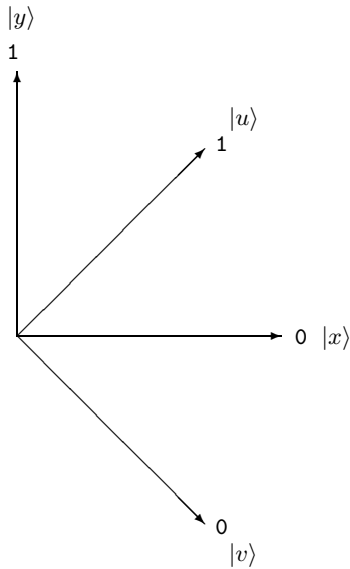
\begin{equation}\label{bried}
P_c=\frac{1}{2}+\frac{1}{4}[\cos 2\theta + \sin 2\theta]
\end{equation}
To find the measurement basis that gives the best chance of
guessing the bit value correctly we maximize $P_c$ with respect to
$\theta$ i.e.
\begin{equation}
\frac{dP_c}{d\theta}=-\frac{1}{2}[\sin2\theta - \cos2\theta]=0
\end{equation}
This gives $\tan 2\theta=1$ or $\theta=\pi/8=22.5^{o}$. $\ket{1}$
is orthogonal to $\ket{0}$ and $22.5^{o}$ clockwise from
$\ket{y}$. This is known as the intermediate or Breidbart basis.
Substituting the value of $\theta$ into equation~(\ref{bried})
gives $P_c=1/2(1+1/\sqrt{2})$. The intermediate states, are now
given by
\begin{equation}
\ket{0}=\cos(\pi/8)\ket{x}-\sin(\pi/8)\ket{y}
\end{equation}
\begin{equation}
\ket{1}=\sin(\pi/8)\ket{x}+\cos(\pi/8)\ket{y}
\end{equation}

We have now found the best basis in which to guess the value of
the bits Alice has encoded on to the quanta. Alice and Bob check
for Eve's presence by measuring the proportion of errors in their
sifted key after the public announcement of the basis. It turns
out that Eve minimizes the disturbance to the sifted key by
recording the value of her measurement, either $M_0$ or $M_1$,
then sending on to Bob one of two states in the measurement basis
either $\ket{0}$ or $\ket{1}$ i.e. if Alice sends $\ket{x}$ Eve
will make her measurement and get the result $M_0$ a proportion
$P_c$ of the time and send Bob $\ket{0}$. Eve may also get the
result $M_1$ a proportion $1-P_c$ of the time and guess the value
of the state incorrectly. In this case she sends $\ket{1}$. Eve
causes an error in Alice and Bob's key when Bob gets the result
that $\ket{y}$ was sent. This probability, $D$, that Eve causes a
disturbance or error is
\begin{equation}\label{dist}
D=P_c\langle 0|M_y|0 \rangle+(1-P_c)\langle 1|M_y|1 \rangle
\end{equation}
Where $M_y=\ket{y}\bra{y}$ is the outcome of Bob's measurement. It
is sufficient to analyze only this case only without the prior
probability weightings of $\ket{x}$, $\ket{y}$ etc. being sent if
they are all equal. Equation~(\ref{dist}) reduces to
\begin{displaymath}
D=2P_c(1-P_c)
\end{displaymath}
\begin{equation}
D=\frac{1}{4}
\end{equation}
If Alice and Bob see that a quarter of the bits in their sifted
key are incorrect while error checking, this is a give away that
Eve was on the line. To mask her presence more carefully Eve can
measure a proportion of the states $\epsilon$. In this case Eve's
probability of guessing the state correctly is
\begin{displaymath}
P_c(\epsilon)=\frac{\epsilon}{2}(1+\frac{1}{\sqrt{2}})+
\frac{1-\epsilon}{2}
\end{displaymath}
Where the last term on the right comes from the fact that Eve can
guess with an 50/50 chance the identity of the bit values she did
not measure. The disturbance becomes
\begin{displaymath}
D(\epsilon)=\frac{\epsilon}{4}
\end{displaymath}
Combining these two expressions we get the probability of Eve
guessing the bit value correctly as a function of $D$.
\begin{equation}
P_c(D)=\sqrt{2}D+\frac{1}{2}
\end{equation}
Note that $D_{max}=1/4$ as this is when $\epsilon=1$.

\subsection{The optimal incoherent strategy}\label{OI}

This strategy is reported from Fuchs {\em et al.} \cite{Fuchs} who
proved it to be the optimum incoherent strategy an eavesdropper
can use on an ideal quantum cryptography system using the BB84
protocol. The derivation given here was first given by Cirac and
Gisin \cite{Cirac} who found a simple symmetry argument that gives
the same result. This attack is not available to an eavesdropper
with present technology.

The idea is to evolve a probe in some initial standard state,
$\ket{a}$, into distinct probe states dependant on the state of
Alice's photon with which the probe is interacted with. The probes
can then be stored until Alice and Bob announce the basis so that
Eve can increase her chance of distinguishing the probe (and
Alice's state) by choosing the best measurement for that
particular basis. Simultaneously Eve wants to make the state she
is sending to Bob as much alike to the state Alice sent to
minimize the disturbance or error rate to the sifted key. We have
the following interactions for the $x$-$y$ basis.
\begin{equation}\label{thestates}
\ket{a}\otimes\ket{x}\longrightarrow
\ket{a^{xy}_{00}}\otimes\ket{x} + \ket{a^{xy}_{01}}\otimes\ket{y}
\end{equation}
\begin{equation}\label{thestates1}
\ket{a}\otimes\ket{y}\longrightarrow
\ket{a^{xy}_{10}}\otimes\ket{x} + \ket{a^{xy}_{11}}\otimes\ket{y}
\end{equation}
with similar interactions for the $u$-$v$ basis. The second kets
in each tensor product on the right hand side after tracing out
the probe states form the mixed state Bob will receive. The first
kets in each tensor product after tracing out Bob's system will
form the mixed state Eve will keep. Notice that the probe states
$\ket{a^{xy,uv}_{ij}}$ are not normalized yet. These two
interactions can be written more neatly in matrix form.
\begin{equation}
\ket{a}\otimes\begin{pmatrix}
  {\ket{x}} \\
  {\ket{y}}
\end{pmatrix} \longrightarrow \varepsilon^{xy}\otimes\begin{pmatrix}
  {\ket{x}} \\
  {\ket{y}}
\end{pmatrix}
\end{equation}
Where $\varepsilon^{xy}$ is the matrix containing the
(un-normalized) probe states. A similar expression can be
constructed for the $u$-$v$ basis simply by changing subscripts.
The probe state matrices in either basis have the form
\begin{equation}
\varepsilon=\begin{pmatrix}
  {\ket{a_{00}}} & {\ket{a_{01}}} \\
  {\ket{a_{10}}} & {\ket{a_{11}}}
\end{pmatrix}
\end{equation}
Using the Hadamard transform, $H$, we can transform this matrix of
probe states from one basis to the other. The transform is
\begin{equation}\label{transform}
\varepsilon^{uv}=H\varepsilon^{xy}H^{\dag}
\end{equation}
where $H$ is
\begin{equation}
H=H^{\dag}=\frac{1}{\sqrt2}\begin{pmatrix}
  {1} & {1} \\
  {1} & {-1}
\end{pmatrix}
\end{equation}
Here we introduce the symmetry part of the argument.
We are going to impose the conditions that\\
(i) All overlaps between probe states should be invariant under
the change of labels $0\longleftrightarrow1$. This means $\langle
a_{10}|a_{11} \rangle=\langle a_{01}|a_{00}\rangle$ etc.\\
(ii) Overlaps between states in one basis should be the same as
overlaps in the other basis. For example $\langle
a^{xy}_{00}|a^{xy}_{00} \rangle=\langle
a^{uv}_{00}|a^{uv}_{00}\rangle$.\\
Going back to equation~(\ref{thestates}) we need to normalize the
probe states and enforce the unitary evolution condition (that
overlaps should be preserved). Normalization requires
\begin{equation}
\langle a_{00}|a_{00} \rangle+\langle a_{01}|a_{01} \rangle=1
\end{equation}
\begin{equation}
\langle a_{11}|a_{11} \rangle+\langle a_{10}|a_{10} \rangle=1
\end{equation}
From the symmetry requirement (i) and defining two parameters $F$
and $D$ we have
\begin{equation}\label{F}
\langle a_{00}|a_{00} \rangle=\langle a_{11}|a_{11} \rangle=F
\end{equation}
and
\begin{equation}\label{D}
\langle a_{01}|a_{01} \rangle=\langle a_{10}|a_{10} \rangle=D
\end{equation}
where $F+D=1$. From the unitary condition i.e. taking the overlaps
of equations~(\ref{thestates}), (\ref{thestates1}) we have
\begin{equation}\label{unitary}
\langle a_{00}|a_{10} \rangle+\langle a_{01}|a_{11} \rangle=0
\end{equation}
and by imposing symmetry condition (i) equation~(\ref{unitary})
becomes
\begin{equation}
\langle a_{11}|a_{01} \rangle+\langle a_{10}|a_{00} \rangle=0
\end{equation}
We shall also impose that all overlaps are real numbers i.e.
$\langle a_{00}|a_{01} \rangle=\langle a_{01}|a_{00} \rangle$
which you can do by careful choices of the local phase of the
states. Now because the overlaps are real
\begin{displaymath}
\langle a_{00}|a_{10} \rangle=0
\end{displaymath}
\begin{displaymath}
\langle a_{01}|a_{11} \rangle=0
\end{displaymath}
\begin{displaymath}
\langle a_{11}|a_{01} \rangle=0
\end{displaymath}
\begin{equation}\label{zeros}
\langle a_{10}|a_{00} \rangle=0
\end{equation}
This means all these probe states are orthogonal to one another.
We will also define some final parameters $F_1$ and $D_1$ that
characterize the overlap of the remainder.
\begin{equation}\label{F1}
\langle a_{00}|a_{11} \rangle=\langle a_{11}|a_{00} \rangle=F_1
\end{equation}
\begin{equation}\label{D1}
\langle a_{10}|a_{01} \rangle=\langle a_{01}|a_{10} \rangle=D_1
\end{equation}
What we are doing is basically defining the unitary transform $U$
that we want to perform on the states. We already know the
relation $F+D=1$, now we need another formula to relate the final
variables $F_1$ and $D_1$. We can do this from symmetry argument
(ii). By transforming the probe states in one basis to their
representation in the other basis using equation~(\ref{transform})
then we get
\begin{equation}
F-D=F_1+D_1
\end{equation}
We can re-write equations~(\ref{thestates}) and (\ref{thestates1})
with the probe states normalized. Normalized probe states from now
on will be denoted with a hat i.e. $\ket{\hat{a}}$. The normalized
states are
\begin{equation}\label{normalizedstates}
\ket{a}\otimes\ket{x}\longrightarrow
\sqrt{F}\ket{\hat{a}^{xy}_{00}}\otimes\ket{x} +
\sqrt{D}\ket{\hat{a}^{xy}_{01}}\otimes\ket{y}
\end{equation}
\begin{equation}\label{normalizedstates2}
\ket{a}\otimes\ket{y}\longrightarrow
\sqrt{F}\ket{\hat{a}^{xy}_{11}}\otimes\ket{y} +
\sqrt{D}\ket{\hat{a}^{xy}_{10}}\otimes\ket{x}
\end{equation}
There are similar relations for the $u$-$v$ basis as well thanks
to the symmetry condition (ii), but because of this symmetry it is
sufficient to analyze just one basis and carry the results over to
the other basis. $F$ as it appears here is the fidelity. Note from
equations~(\ref{normalizedstates}), (\ref{normalizedstates2}) that
these are entangled states. Looking specifically at
equation~(\ref{normalizedstates}), when Bob makes his measurement
in the basis that Alice sent the state in, the only states that
will eventually form the sifted key, he can get two outcomes.
Either the state was $\ket{x}$ which occurs with probability $F$
in which case the superposition will collapse and Eve's probe
state instantaneously jumps to $\ket{\hat{a}^{xy}_{00}}$. This
means Eve has got away without causing a disturbance. The other
outcome is $\ket{y}$, an error or disturbance in Alice and Bob's
sifted key occurring with probability $D$. Just because Eve got
away without causing disturbance a proportion $F$ of the time does
not necessarily mean she will conclude correctly what the state
Alice sent was. Now Eve has to discriminate between two density
matrices because she can store her probes until the basis is
publicly announced and choose the best measurement to determine
which density matrix it was. Eve's mixed states are
\begin{equation}
\rho_i=tr_{Bob}[U(\ket{a}\otimes\ket{i}\bra{a}\otimes\bra{i})U^{\dag}]
\end{equation}
The density matrices describing Eve's probes after the $x$-$y$
basis has been announced are
\begin{equation}
\rho_x=F\ket{\hat{a}_{00}}\bra{\hat{a}_{00}} +
D\ket{\hat{a}_{01}}\bra{\hat{a}_{01}}
\end{equation}
\begin{equation}
\rho_y=F\ket{\hat{a}_{11}}\bra{\hat{a}_{11}} +
D\ket{\hat{a}_{10}}\bra{\hat{a}_{10}}
\end{equation}
From equations~(\ref{zeros}) we can see the four probe states form
two orthogonal sets. One set
$\{\ket{\hat{a}_{10}},\ket{\hat{a}_{01}}\}$ occurs with
probability $D$ and the other set
$\{\ket{\hat{a}_{00}},\ket{\hat{a}_{11}}\}$ occurs with
probability $F$. What Eve can do is devise a measurement that
unambiguously discriminates these two sets. Because the
measurement will also be compatible with the sets it will not
change the states of the probes. Eve can also tell, depending on
the result of this first measurement, whether she has caused a
disturbance to Alice and Bob's key. The second problem is now to
discriminate between two, generally nonorthogonal states in the
same set. The basis that gives Eve the best chance of guessing the
state correctly is the intermediate or Breidbart basis. For two
states $\ket{\psi_1}$ and $\ket{\psi_2}$ the probability of
guessing the state correctly is \cite{Helstrom}
\begin{equation}
P_c=\frac{1}{2}+\frac{1}{2}\sqrt{1-|\langle
\psi_1|\psi_2\rangle|^2}
\end{equation}
We know the overlaps between the two sets from
equations~(\ref{F}), (\ref{D}), (\ref{F1}) and (\ref{D1}). These
are
\begin{equation}
\langle \hat{a}_{00}|\hat{a}_{11}\rangle=\langle
\hat{a}_{11}|\hat{a}_{00}\rangle=\frac{F_1}{F}
\end{equation}
\begin{equation}
\langle \hat{a}_{01}|\hat{a}_{10}\rangle=\langle
\hat{a}_{10}|\hat{a}_{01}\rangle=\frac{D_1}{D}
\end{equation}
So the probability of Eve guessing the correct state is
\begin{equation}\label{naaaa}
P_c=FP^{F}_c+DP^{D}_c
\end{equation}
Where $P^{F}_c=1/2+1/2\sqrt{1-(F_1/F)^2}$ for example. It actually
turns out that equation~(\ref{naaaa}) is maximized when the probes
in both sets have the same overlap i.e.
\begin{equation}\label{naa2}
\frac{F_1}{F}=\frac{D_1}{D}
\end{equation}
So $P^{F}_c=P^{D}_c$ and because $F+D=1$ we have the optimum
incoherent eavesdropping strategy. Eve has the probability to
guess correctly
\begin{equation}
P_c=\frac{1}{2}+\frac{1}{2}\sqrt{1-(\frac{D_1}{D})^2}
\end{equation}
Using the relations $F+D=1$, $F-D=F_1+D_1$ and
equation~(\ref{naa2}) we can express $P_c$ in terms of $D$ only
giving
\begin{equation}\label{theoptprob}
P_c=\frac{1}{2}+\sqrt{D(1-D)}
\end{equation}
This strategy does much better than the intercept-resend strategy
at the cost of added technological complexity. Griffiths and Niu
\cite{Griffiths} have proposed the quantum circuit that does the
job. This strategy can also produce deterministic information on
every state Alice sends to Bob at the price of completely
randomizing their sifted key. This is a good example of the
information-disturbance trade off. Eve has a better chance of
distinguishing her probe states the smaller she makes their
overlap until they become orthogonal to one another and can be
discriminated perfectly. The price for decreasing the overlap
between Eve's probes is that Bob's density matrices become more
uniformly mixed between the wanted and the unwanted states i.e.
$D$ increases. This comes from nature's enforcement of unitary
evolution, that all the overlaps must remain the same before and
after the evolution.

We know this is the optimum strategy from the derivation given by
Fuchs {\em et al.} \cite{Fuchs}. The other interesting thing about
this strategy is that can be regarded as an asymmetrical cloning
machine, taking Alice's state and producing two copies of unequal
fidelity. One good copy is passed on to Bob and one bad copy that
is kept by Eve. There are bounds on copying machines and the
result in equation~(\ref{theoptprob}) can be found from work by
Cerf {\em et al.} \cite{Cerf}. However in the derivation presented
above the probe lives in a $4$ dimensional Hilbert space, in other
words the probe consists of two qubits. Using a cloning machine
the probe need only be of $2$ dimensions, a single qubit. This has
been shown by Niu and Griffiths \cite{NiuQcopying}.

\subsection{What can Alice and Bob do?}\label{countermeasures}
What can Alice and Bob do about this situation if they know about
both of these strategies? The condition is that Alice and Bob can
recover a secret key by one way privacy amplification if Eve's
mutual information on Alice or Bob's bit string is less than the
mutual information Alice and Bob share \cite{EkertEve, Csiszar}.
Formally this constraint is $I(A;B)\geq max\{I(E;B),I(A;E)\}$,
where $I(A;B)$ is the mutual information between Alice and Bob,
$I(A;E)$ is the mutual information between Alice and Eve and
$I(E;B)$ is the mutual information Eve and Bob share. We use the
one way result for simplicity.

\begin{figure}
\centering
\includegraphics[width=8cm]{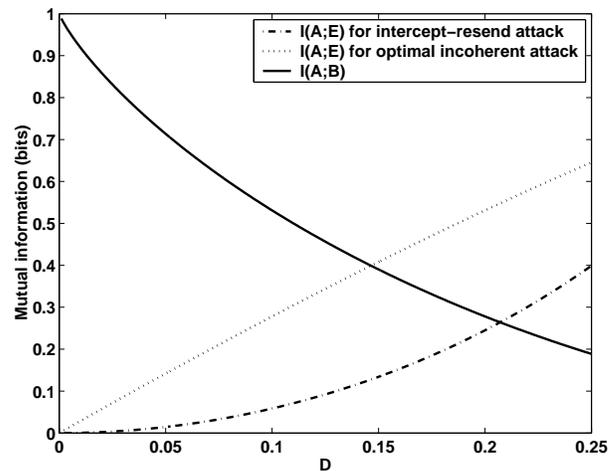}
\caption{The average mutual information as a function of the error
rate, $D$, for each strategy. The point at which $I(A;B)$ crosses
with $I(A;E)$ is the point at which Alice and Bob can no longer
distill a secret key by one way privacy
amplification.}\label{singlephotonfig}
\end{figure}

The average mutual information Alice and Bob share on the sifted
key after public announcement of the basis in the presence of an
eavesdropper is just the average mutual information on a binary
symmetric channel \cite{Cover}. The disturbance, $D$, an
eavesdropper introduces looks like a data flipping rate to Alice
and Bob.
\begin{equation}\label{binarychannel}
I(A;B)=\log 2+D\log D+(1-D)\log(1-D)
\end{equation}
$I(A;B)$ can be expressed in bits by taking the logarithms to the
base $2$ or nats by taking logarithms to the base $e$.
Equation~(\ref{binarychannel}) can be expressed more symmetrically
in terms of a function $\phi(z)$ \cite{Fuchs} as
\begin{equation}\label{one}
I(A;B)=\frac{1}{2}\phi(1-2D)
\end{equation}
Where $\phi(z)$ is
\begin{equation}
\phi(z)=(1-z)\log(1-z)+(1+z)\log(1+z)
\end{equation}
The average mutual information Alice and Eve, $I(A;E)$, and Eve
and Bob share, $I(E;B)$, are equal for these strategies. The
probability of the bit being flipped in this case is just $1-P_c$.
Expressed in terms of the function $\phi(z)$ this is
\begin{equation}\label{two}
I(A;E)=I(E;B)=\frac{1}{2}\phi[1-2(1-P_c)]
\end{equation}
By comparing equation~(\ref{one}) with equation~(\ref{two}) we can
find the maximum disturbance at which Alice and Bob can expect to
obtain a secret key by privacy amplification i.e. $I(A;B)>I(A;E)$.
For a secret key to be obtained by \emph{one way} privacy
amplification
\begin{equation}
D< 1-P_c
\end{equation}
For the intercept-resend strategy this is
\begin{equation}
D<\frac{1}{2(1+\sqrt{2})}
\end{equation}
for secret key generation to be possible and for the optimal
incoherent strategy
\begin{equation}
D<\frac{2-\sqrt{2}}{4}
\end{equation}
This means if the channel has a noise level greater than this
value then Alice and Bob cannot use it for quantum key generation
as they should assume all disturbance to derive from an
eavesdropper \cite{Fuchs} if they are using one way privacy
amplification. If they detect disturbance above this level then
they must try the whole key generation process again or give up.
However, in a practical implementation of quantum cryptography
this may be over-conservative \cite{Felix}. The average mutual
information between Alice and Eve and Alice and Bob in bits for
both strategies is shown in figure~\ref{singlephotonfig}.

\section{Incoherent eavesdropping: laser pulses}\label{ars}

In this section we use the results from section~\ref{ais} to form
better strategies when the light source is a laser pulse that can
be modelled by a mixture of number states with a Poissonian photon
number distribution. All of these attacks use a beam-splitter
apart from the photon-number splitting attack. In the
photon-number splitting attack an eavesdropper can make a decision
on what her best action should be because she can measure the
number of photons in the pulse non-destructively. We start by
introducing a few tools needed for this next part.

\subsection{The beam-splitter}

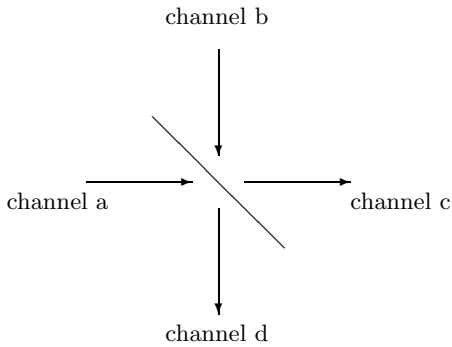
\begin{figure}
\centering
\begin{picture}(200,200)
\put(75,125){\line(1,-1){50}}
\put(20,90){channel a}
\put(150,90){channel c}
\put(80,40){channel d}
\put(80,160){channel b}
\put(100,90){\vector(0,-1){40}}
\put(50,100){\vector(1,0){40}}
\put(110,100){\vector(1,0){40}}
\put(100,150){\vector(0,-1){40}}
\end{picture}
\caption{A beam-splitter: Photons going into channel c have been
transmitted with probability $t$, photons going into channel d
have been reflected with probability $r$ if all photons arrive
from channel a.}\label{bs}
\end{figure}

A beam-splitter either reflects a photon with some probability $r$
or it transmits it with probability $t$. We are going to be
dealing with lossless beam-splitters so $r+t=1$. The beam-splitter
is shown in figure~\ref{bs}.
We can describe the action of the beam-splitter on a number state
$\ket{n}$ entering through channel a and the vacuum state
$\ket{0}$, entering through channel b as
\begin{equation}
\ket{n}_a\otimes\ket{0}_b\longrightarrow\sum_{i=0}^{n}c_i\ket{i}_c\otimes\ket{n-i}_d
\end{equation}
The $c_i$'s are the probability weighting for each situation. If
we take the reflection or transmission of a photon as being
independent of each other than the $c_i$ are given by the binomial
formula normalized appropriately
\begin{equation}
c_i=[\begin{pmatrix}
  {n} \\
  {i}
\end{pmatrix}t^{i}r^{n-i}]^{\frac{1}{2}}
\end{equation}
Where
\begin{displaymath}
\begin{pmatrix}
  {n} \\
  {i}
\end{pmatrix}=\frac{n!}{i!(n-i)!}
\end{displaymath}
So the probability of detecting $j$ photons in channel c is
\begin{displaymath}
|(\bra{j}_c\otimes\bra{n-j}_d)(\sum_{i=0}^{n}c_i\ket{i}_c\otimes\ket{n-i}_d)|^2
\end{displaymath}
\begin{equation}
=\frac{n!}{j!(n-j)!}t^{j}(1-t)^{n-j}
\end{equation}

\subsection{Representation of laser pulses}

In the ideal situation one would what to use a light source that
emits single photons on demand. Unfortunately sources that emit
single photon number states are not available yet. Practical
quantum cryptography usually uses an attenuated laser pulse. These
pulses can be modelled as a mixed state with a Poisson number
state distribution or as a superposition of number states with a
Poisson distribution, a coherent state. In the following analysis
we calculate values using a coherent state as the input although
strictly speaking the input state is mixed when the logical bit
values are encoded onto the polarisation. However, whichever input
state is used gives the same answers for the values we are
interested in, those given in section~\ref{scenarios} and onwards.
The probability of detecting $n$ photons in a pulse with mean
photon number $\mu$ is given by
\begin{equation}\label{probnumberstate}
p_n=e^{-\mu}\frac{\mu^n}{n!}
\end{equation}
If Alice and Bob's quantum channel, the channel they use for
sending the states that make up the key is lossy, the probability
of detecting $n$ photons when the pulse gets to Bob
is
\begin{equation}\label{loss}
p_n=e^{-\eta\mu}\frac{(\eta\mu)^n}{n!}
\end{equation}
Where $\eta$ is the transmission efficiency of the channel, the
probability that each photon has of being detected by Bob. When
$\eta=1$ the channel has no loss, all photons in the pulse make it
to Bob. When $\eta=0$ the channel is completely opaque.

The operation of the beam-splitter on a light pulse entering from
channel a (figure~\ref{bs}) looks like
\begin{equation}
\sum_{n=0}^\infty \sqrt{p_n}
\ket{n}_a\otimes\ket{0}_b\longrightarrow\sum_{n=0}^\infty
\sqrt{p_n} \sum_{i=0}^{n}c_i\ket{i}_c\otimes\ket{n-i}_d
\end{equation}
The probability to find $j$ photons in channel c is
\begin{equation}\label{dist1}
|(\sum_{m=0}^{\infty}\bra{j}_c\otimes\bra{m-j}_d)(\sum_{n=0}^\infty
\sqrt{p_n} \sum_{i=0}^{n}c_i\ket{i}_c\otimes\ket{n-i}_d)|^2
\end{equation}

\subsection{Eavesdropping attacks using a beam-splitter}\label{scenarios}

We can know formulate an incoherent attack based on
beam-splitting. In a practical implementation, a laser produces
states that give the chance that multiple photons will be
detected, a good scenario for an eavesdropper trying to evade
detection. All current quantum cryptographic systems use these
sources and this will be the case until reliable, efficient,
single photon sources become available. Work has also been done on
these attacks by \cite{DusekOC, Felix}.

The eavesdropper has positioned her beam-splitter in Alice and
Bob's quantum channel and replaced their lossy quantum channel
with a lossless one ($\eta=1$). The beam-splitter is characterized
by its transmission coefficient, $t$, and its reflection
coefficient $r=1-t$. These are just the probabilities that a
photon incident on the beam-splitter is transmitted or reflected
respectively. Alice sends polarisation encoded Poissonian
distribution mixed photon number states to Bob. These are
characterized by their mean photon number $\mu$. We have four
scenarios that can happen during Eve's attack labelled A, B, C and
D. These scenarios will be quoted when formulating specific
attacks later.

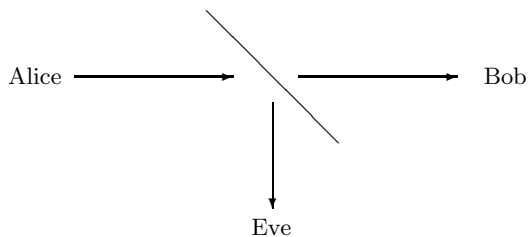
\begin{figure}
\centering
\begin{picture}(200,120)(0,40)
\put(75,125){\line(1,-1){50}}
\put(0,97){Alice}
\put(180,97){Bob}
\put(92,40){Eve}
\put(100,90){\vector(0,-1){40}}
\put(25,100){\vector(1,0){60}}
\put(110,100){\vector(1,0){60}}
\end{picture}
\caption{Beam-splitter attack scenario A. The arrows in Alice's,
Bob's and Eve's channels represent the detection of at least one
photon. The eavesdropper profits most from this situation.}
\label{scenarioA}
\end{figure}

\subsubsection{Scenario A (figure~\ref{scenarioA})}
Alice has emitted a state with more than one photon in it.
\emph{At least} one of these photons is reflected at the
beam-splitter and travels into Eve's channel \emph{and at least}
one photon is transmitted by the beam-splitter into Bob's channel.
This means Eve can obtain this state for free. She can store the
photon provided she has the technology, until Alice and Bob
announce the basis publicly and then discover the photon's state
unambiguously. In these circumstances Eve does not cause an error
on the sifted key.

Scenario A is the ideal outcome for Eve, she cannot be detected
and she gets deterministic information provided she can store and
preserve the photon or probe's state until public announcement of
the basis.
For this scenario to occur there needs to be at least two photons
and at least one in each channel. To find the probability for
scenario A to occur this means we have to alter the summation
limits for the beam-splitter operation on the pulse to give
\begin{equation}
P_A=\sum_{n=2}^{\infty}p_n\sum_{i=1}^{n-1}|c_i|^2
\end{equation}
We can replace the beam-splitter part, the last summation on the
right by reordering i.e.
\begin{equation}
\sum_{i=1}^{n-1}|c_i|^2=1-|c_0|^2-|c_n|^2=1-r^n-t^n
\end{equation}
Putting in the explicit form of $p_n$ and making use of $r=1-t$ we
have
\begin{equation}\label{a99}
P_A=e^{-\mu}\sum_{n=2}^{\infty}\frac{\mu^n}{n!}[1-(1-t)^n-t^n]
\end{equation}
The remaining summation can also be dealt with using the same
trick
\begin{equation}
\sum_{n=0}^{\infty}\frac{x^n}{n!}=1+x+\frac{x^2}{2!}+.....+\frac{x^n}{n!}=e^x
\end{equation}
So with the limits in equation~(\ref{a99})
\begin{equation}
\sum_{n=2}^{\infty}\frac{x^n}{n!}=e^x-x-1
\end{equation}
After some algebra $P_A$ is
\begin{displaymath}
P_A=1+e^{-\mu}-e^{-\mu t}-e^{-\mu(1-t)}
\end{displaymath}
\begin{equation}
=1+e^{-\mu}-2e^{-\frac{\mu}{2}}\cosh[\mu(t-1/2)]
\end{equation}
$P_A$ is maximized when $t=1/2$ (i.e. $\cosh(x)$ reaches a minimum
when $x=0$) but maximizing this probability means changing the
probabilities for the other scenarios which may not be the best
thing to do as Eve is trying to make her presence as inconspicuous
as possible.

\subsubsection{Scenario B (figure~\ref{scenarioB})}
Alice has emitted a state with one or more photons in it. However
all the photons in the state are reflected into Eve's channel.
This is a bad situation for Eve. There is no chance that this
state will form part of Alice and Bob's sifted key and the second
part of the bad news is that Eve is lowering the proportion of the
time Bob is expecting to receive a click at his detectors. Bob
expects a certain proportion of the time not to receive a photon
because sometimes the pulse sent by Alice is the vacuum state and
Alice and Bob's original channel has a transmission efficiency. He
will know what proportion this should be by calculating it from
the settings of Alice's equipment; what she has set the mean
photon number, $\mu$, of the pulses to be by running the light
through an attenuator and by making measurements of the channel
transmission efficiency, $\eta$. This scenario reduces the
proportion of states sent that have at least one photon in them
from Bob's point of view.

\begin{figure}
\centering
\begin{picture}(200,120)(0,40)
\put(75,125){\line(1,-1){50}}
\put(0,97){Alice}
\put(180,97){Bob}
\put(92,40){Eve}
\put(25,100){\vector(1,0){60}}
\put(110,100){\vector(1,0){60}}
\end{picture}
\caption{Beam-splitter attack scenario C. The arrows in Alice's
and Bob's channels represent the presence of at least one photon.}
\label{scenarioC}
\end{figure}
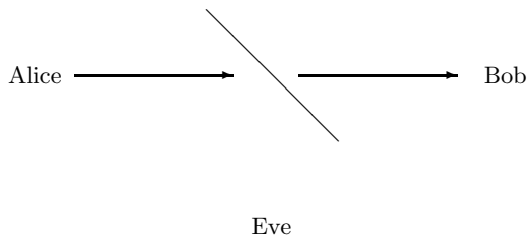

The probability for this occurrence is
\begin{equation}
P_B=\sum_{n=1}^{\infty}p_n|c_0|^2=e^{-\mu}\sum_{n=1}^{\infty}\frac{\mu^n}{n!}(1-t)^n
\end{equation}
Again rearranging the summation we have
\begin{equation}
P_B=e^{-\mu t}-e^{-\mu}
\end{equation}
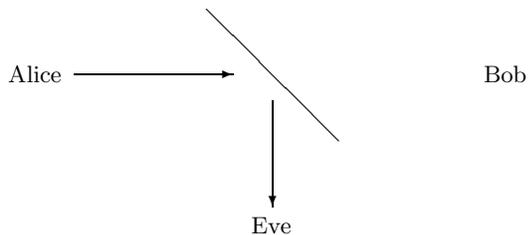
\begin{figure}
\centering
\begin{picture}(200,120)(0,40)
\put(75,125){\line(1,-1){50}}
\put(0,97){Alice}
\put(180,97){Bob}
\put(92,40){Eve}
\put(100,90){\vector(0,-1){40}}
\put(25,100){\vector(1,0){60}}
\end{picture}
\caption{Beam-splitter attack scenario B. The arrows in Alice's
and Eve's channels represent the detection of at least one photon.
Note the absence of a photon in Bob's channel.} \label{scenarioB}
\end{figure}
\subsubsection{Scenario C (figure~\ref{scenarioC})}
In this scenario all photons in the pulse are transmitted by the
beam-splitter into Bob's channel. This is not as bad a situation
as scenario B but it is not as good as scenario A. The states in
this scenario may go on to form part of Alice and Bob's sifted key
and Eve does not cause any additional reduction in the proportion
of non-empty pulses reaching Bob. Just because Eve does not
receive a photon in her channel does not mean she need be
completely ignorant of its identity. Eve could employ one of the
single photon strategies in section~\ref{ais} at the cost of
introducing errors on Alice and Bob's sifted key.
The probability that scenario C occurs (all photons end up in
Bob's channel) is
\begin{equation}
P_c=\sum_{n=1}^{\infty}p_n|c_n|^2=
e^{-\mu}\sum_{n=1}^{\infty}\frac{\mu^n}{n!}t^n
\end{equation}
Which becomes
\begin{equation}
P_c=e^{-\mu(1-t)}-e^{-\mu}
\end{equation}
\subsubsection{Scenario D}
This is a simple occurrence. This is just the situation that the
pulse Alice emits has no photon in it. It is the vacuum state. In
this case neither Alice, Bob or Eve can get any information. The
probability of zero photons in the state is
\begin{equation}
P_0=e^{-\mu}
\end{equation}
\subsubsection{Photon number distribution after the beam-splitter}
When Eve inserts a beam-splitter into Alice and Bob's quantum
channel she alters the probability of the number of photons Bob
detects. The probability that Bob finds $i$ photons in his channel
was given by equation~(\ref{dist1}). Written explicitly this is
\begin{displaymath}
P(i)=e^{-\mu}\frac{(\mu
t)}{i!}^i\sum_{n=i}^{\infty}\frac{[\mu(1-t)]}{(n-i)!}^{n-i}
\end{displaymath}
\begin{equation}
=e^{-\mu t}\frac{(\mu t)^i}{i!}
\end{equation}
On comparison with equation~(\ref{loss}) we can see that the
addition of the beam-splitter does not change the overall shape of
the photon number distribution; it is still Poissonian (although
with a smaller average number of photons). If Eve then replaces
Alice and Bob's lossy channel described by $\eta$ with one that is
lossless ($\eta=1$) she can then choose the transmission
coefficient of the beam-splitter, $t$ to match $\eta$ ($\eta=t$).
In this way Eve does not cause any change the photon statistics
Bob expects.

\subsection{Intercept-resend attack with a beam-splitter}

A strategy like this could be implemented with today's technology.
This section borrows the main results from section~\ref{idealI-R}
and section~\ref{scenarios}.

This is what Eve does: If she gets photons in her channel she
measures them in every instance that the situation occurs. When
scenario A occurs, only the photons in Bob's channel will be
subject to error checking so she will not cause any errors on the
sifted key. Borrowing the main result from section~\ref{idealI-R}
Eve's probability of guessing the state correctly from scenario A
alone is
\begin{equation}
P^{i-r}(D=\frac{1}{4})=\frac{\sqrt{2}+2}{4}
\end{equation}
The overall probability that Eve guesses the correct state when
scenario A occurs is
\begin{equation}
P^{correct}_A=\frac{P_A P^{i-r}(D=1/4)}{1-P_0-P_B}
\end{equation}
The denominator in the equation above appears because we only
include those situations in which the states have the opportunity
to form part of Alice and Bob's sifted key. It is like a
normalizing factor. Eve may get more than one photon in her
channel when scenario A occurs. If she then measures them in the
intermediate basis she may get both measurement results $M_0$ and
$M_1$. What should Eve conclude now? We make an approximation here
and say that Eve always gets a single result. To give some
justification of this if Eve has a detector that can measure the
number of photons incident on it but destroys the photons in the
process then when Eve analyzes more than one photon she can be
reasonably certain that the bit value sent was the one that
resulted in the higher number of photon detections for a that
outcome. The only time Eve must make a 50/50 guess is when there
are the same number of detected photon for each measurement
outcome. However, more than one photon detected for only one
outcome will give Eve a higher probability of guessing the state
correctly so we think this is not a bad approximation.

The second part of the strategy is that when Eve does not get a
photon in her channel she measures a proportion, $\epsilon$, of
the photons that make it into Bob's channel. She only does this a
proportion of the time to control the level of disturbance she
inflicts on Alice and Bob's sifted key. The disturbance, $D_{AB}$,
that Alice and Bob see in their sifted key is going to be
different from the $D$ appearing in the single photon strategies
because they only see this disturbance on a proportion of the
bits, those that came from scenario C namely
\begin{equation}
\frac{\epsilon P_c}{1-P_0-P_B}
\end{equation}
Therefore the disturbance Alice and Bob see is
\begin{equation}\label{DAB}
D_{AB}(\epsilon)=\frac{D(\epsilon)P_C}{1-P_0-P_B}
\end{equation}
The probability of guessing the state correctly from scenario C is
given by
\begin{equation}
P_C^{correct}(D)=\frac{P^{i-r}(D)P_C}{1-P_0-P_B}
\end{equation}
Where (borrowed from section~\ref{idealI-R})
\begin{equation}
P^{i-r}(D)=\sqrt{2}D+\frac{1}{2}
\end{equation}
So the total probability of guessing the state correctly is
\begin{equation}\label{final}
P^{correct}_{tot}=\frac{P_A P^{i-r}(D=1/4)+ P_C
P^{i-r}(D)}{1-P_0-P_B}
\end{equation}
The probabilities of each of the scenarios is dependant on the
transmission coefficient of the beam-splitter, $t$. From
section~\ref{scenarios} it was found that if Alice and Bob are
expecting to see a line with loss $\eta$ then Eve can replace it
with a lossless one and make her beam-splitter transmission be
$t=\eta$ to give the same photon number statistics Bob expects.
Now what Eve has to do to avoid Alice and Bob's suspicions is fix
the value of $t$ to the value of $\eta$ as well as lowering the
error rate Alice and Bob see, $D_{AB}$.

Giving the explicit form of equation~\ref{final} we have
\begin{equation}\label{i-rtot}
P^{correct}_{tot}=\frac{\sqrt{2}+2}{4} -
e^{-\mu(1-t)}[\sqrt{2}(1/4-D)]
\end{equation}
The disturbance Alice and Bob see (the error rate in their sifted
keys) is
\begin{equation}\label{t2}
D_{AB}=De^{-\mu(1-t)}
\end{equation}
With equation~(\ref{t2}) Eve can control the error rate and by
knowing $\eta$ she can then calculate how likely she is to guess
the correct values of the bits using equation~(\ref{i-rtot}).
\subsection{Optimal incoherent attack with a beam-splitter}

This attack is similar to the last one considered except now the
eavesdropper has better technology and can use the strategy
reported in section~\ref{OI} and extract more information for a
fixed disturbance. When scenario A occurs this time, Eve can
perform an interaction between the probe and the photon in her
channel to maximize the disturbance to the photon's state and
maximize the distinguishability of the probe. She can then store
the probe and wait until the basis is announced before measuring
it. Now she can unambiguously determine which state Alice sent to
Bob. When scenario C occurs the eavesdropper reverts to the single
photon strategy and again Alice and Bob will notice a smaller
average disturbance $D_{AB}$ then $D$ because this situation does
not occur all the time. Eve again should set $\eta=t$ to give Bob
the same photon statistics and $D_{AB}$ will be given by
equation~(\ref{t2}).

So scenario A occurs. The eavesdropper has a probability of being
correct from this circumstance that is
\begin{equation}
P_A^{correct}=\frac{P_A}{1-P_0-P_B}
\end{equation}
There is no additional multiplier because Eve gains deterministic
information i.e. $P^{opt}(D=0.5)=1$.

Scenario C occurs giving a probability of being correct
\begin{equation}
P_C^{correct}=\frac{P_C P^{opt}(D)}{1-P_0-P_B}
\end{equation}
$P^{opt}(D)$ is the probability of guessing the state correctly
taken from section~\ref{OI} and is as a reminder
\begin{equation}\label{op}
P^{opt}(D)=\frac{1}{2}+\sqrt{D(1-D)}
\end{equation}
The total probability that Eve guesses the correct state is
\begin{equation}
P^{correct}_{tot}=\frac{P_A+P_C P^{opt}(D)}{1-P_0-P_B}
\end{equation}
The total probability of Eve guessing the state correctly given
explicitly is
\begin{equation}
P^{correct}_{tot}=1-e^{-\mu(1-t)}[1/2-\sqrt{D(1-D)}]
\end{equation}

\subsection{Photon-number splitting attack}

What an eavesdropper can do here is perform unlimited numbers of
quantum non-demolition (QND) measurements with unit efficiency.
QND measurements are measurements that give the value of one
observable without effecting the other degrees of freedom. The QND
measurement Eve is interested in is one which tells her the number
of photons in the laser pulse without effecting their
polarisation. These measurements are very difficult to perform,
the first demonstration being by Nogues {\em et al.}
\cite{QNDnature}. This eavesdropping attack is far beyond present
technology. Work on these kinds of attack has also been done by
\cite{Lutkenhaus, Brassard}. The flip side is that if Alice has
the technology to perform QND measurements then she can simply
stop any pulse that has more than one photon in it before sending
it to Bob. In this case an eavesdropper will have revert to the
strategies given in section~\ref{ais}. We will here analyze the
case of the more advanced eavesdropper.

Using this attack Eve can discover the number of photons in the
pulse and then choose her actions accordingly. If there is more
than one photon in the pulse she just splits off a single photon
and lets the others pass to Bob. She can then perform an
interaction with her probe and store it until the basis is
announced then make a measurement that will give her the correct
value every time. If there is more than one photon in the pulse
then Eve does not cause an error on Alice and Bob's sifted key. If
she sees only one photon then she performs the optimal incoherent
attack at the cost of adding some disturbance. If Alice and Bob
have a lossy channel with some transmission efficiency $\eta$, Eve
can replace this channel with a lossless one then selectively
block the single photon pulses to give the same proportion of
non-empty pulses Bob is expecting. In this way Eve may even be
able to obtain deterministic knowledge of Alice and Bob's sifted
key without causing \emph{any} errors depending on the values of
the original channel transmission efficiency $\eta$ and the mean
photon number $\mu$.

From equation~(\ref{probnumberstate}) the probability that Alice
sends a pulse containing zero and one photons are
\begin{equation}
P_0=e^{-\mu}
\end{equation}
\begin{equation}
P_1=\mu e^{-\mu}
\end{equation}
And the probability for detecting more than one photon is
\begin{equation}
P_{n>1}=1-e^{-\mu}(1+\mu)
\end{equation}
If Eve blocks a proportion $\kappa$ of the pulses containing one
photon then her probability of guessing the state correctly is
\begin{equation}
P^{correct}_{tot}(D)=\frac{P_{n>1}+
(1-\kappa)P_1P^{opt}(D)}{1-P_0-\kappa P_1}
\end{equation}
The denominator is again there to exclude the circumstances that
Alice and Bob (and Eve in this case) do not use, all the empty
pulses. The probability of Eve guessing the state correctly is
given explicitly as
\begin{widetext}
\begin{equation}
P^{correct}_{tot}(D)=\frac{1-e^{-\mu}(1+\mu)+(1-\kappa)\mu
e^{-\mu}[1/2+\sqrt{D(1-D)}]}{1-e^{-\mu}(1+\mu\kappa)}
\end{equation}
\end{widetext}
The disturbance Alice and Bob see when Eve uses this strategy is
\begin{displaymath}
D_{AB}=\frac{(1-\kappa)P_1}{1-P_0-\kappa P_1}D
\end{displaymath}
\begin{equation}
=\frac{(1-\kappa)\mu e^{-\mu}}{1-e^{-\mu}(1+\mu\kappa)}D
\end{equation}
If Alice and Bob's quantum channel has transmission efficiency
$\eta$, Eve can block a proportion, $\kappa$, of the one photon
pulses if she replaces Alice and Bob's channel with a lossless one
($\eta=1$). We can equate the number of non-empty pulses Bob
expects to see with a lossy channel to the number of non-empty
pulses after Eve selectively blocks the single photon pulses.
\begin{equation}
1-e^{-\eta\mu}=(1-\kappa)P_1+P_{n>1}
\end{equation}
This gives $\kappa$ as
\begin{equation}
\kappa=\frac{1}{\mu}[e^{\mu(1-\eta)}-1]
\end{equation}
When $\kappa=1$ Eve has deterministic knowledge of the
\emph{whole} of Alice and Bob's key without causing \emph{any}
errors in it. Lossy lines and high mean photon numbers are
therefore a huge security risk under this attack. The value of
$\eta$ as a function of the mean photon number, $\mu$, this occurs
at is given by
\begin{equation}
\eta=1-\frac{1}{\mu}\ln(1+\mu)
\end{equation}
\subsection{What can Alice and Bob do?}\label{countermeasures2}

\begin{figure}
\centering
\includegraphics[width=8cm]{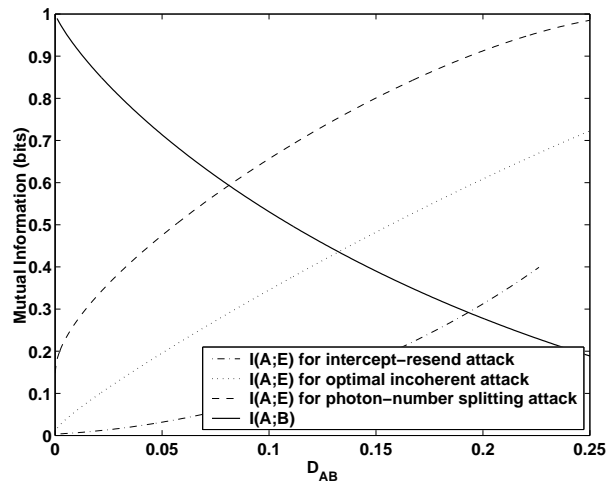}
\caption{The average mutual information as a function of the error
rate, $D_{AB}$, for each strategy at a fixed $\eta=0.9$ and
$\mu=1$. The point at which $I(A;B)$ crosses with $I(A;E)$ is the
point at which Alice and Bob can no longer distill a secret key by
one way privacy amplification.}\label{faintpulsefig}
\end{figure}

If Alice and Bob are trying quantum key distribution with
attenuated laser pulses instead of a single photon source what
measures can they take to ensure that Eve does not obtain their
key? The first thing Alice and Bob can do is check the error rate
in their keys. Using the same criteria as in
section~\ref{countermeasures} for one way privacy amplification to
be successful then
\begin{displaymath}
D_{AB}< 1-P^{correct}_{tot}
\end{displaymath}
If Eve is using the intercept-resend strategy presented here Alice
and Bob must detect a lower disturbance than
\begin{equation}
D_{AB}<\frac{2-\sqrt{2}(1-e^{-\mu(1-\eta)})}{4(1+\sqrt{2})}
\end{equation}
to be able to distill a secret key. If Eve is using the optimal
incoherent attack then
\begin{equation}
D_{AB}<(\frac{2-\sqrt{2}}{4})e^{-\mu(1-\eta)}
\end{equation}
And finally if Eve is using the photon-number splitting attack
then
\begin{equation}
D_{AB}<(\frac{2-\sqrt{2}}{4})[\frac{(1+\mu)e^{-\mu}-e^{-\eta\mu}}{1-e^{-\eta\mu}}]
\end{equation}
The average mutual information as a function of $D_{AB}$, the
error rate in Alice and Bob's sifted key is plotted for each of
these strategies in figure~\ref{faintpulsefig} for $\eta=0.9$ and
$\mu=1$.

Apart from measuring the disturbance or error rate in their keys,
Alice and Bob can also monitor coincidence counts \cite{Felix}.
Bob can count the number of times both of his detectors click when
he chooses the wrong basis i.e. he gets both results from his
measurement. These occurrences come from multiple photons in the
laser pulse. Bob knows the mean photon number of the pulses Alice
is sending and the transmission efficiency of their channel so he
can work out how often this should occur. This is what he can do
to monitor the photon number distribution if he does not have a
detector that can discriminate the number of photons incident on
his detector. Bob expects
\begin{equation}\label{coin}
P_{coincidence}=\frac{1}{2}e^{-\eta\mu}\sum_{n=2}^{\infty}
\frac{(\eta\mu)^n}{n!}\sum_{i=1}^{n-1}\begin{pmatrix}
  n \\
  i
\end{pmatrix}2^{-n}
\end{equation}
\begin{equation}
P_{coincidence}=\frac{1}{2}(1+e^{-\eta\mu}-2e^{-\frac{\eta\mu}{2}})
\end{equation}
The $2^{-n}$ term on the right hand side of equation~(\ref{coin})
is because when Bob uses the wrong basis he effectively has a
50/50 beam-splitter and the $1/2$ factor comes from the fact that
Bob chooses the wrong basis half the time. Eve will not effect the
photon number distribution that Bob sees in the beam-splitting
attacks providing she has set the beam-splitter transmission to
equal the transmission efficiency, $\eta$, of their original lossy
quantum channel as shown in section~\ref{scenarios}. The photon
number distribution is still Poissonian. However, with the more
powerful QND attack which is dependant on the number of photons in
the pulse the photon number statistics Bob sees will no longer be
Poissonian. In the strategy above we fixed the number of non-empty
pulses to be equivalent to Bob's expectations. In doing this
though we have reduced the probability of Bob finding a pulse
containing more than one photon which Bob can detect from the
reduction in the proportion of coincidence counts. A more detailed
analysis of this is given by F\'{e}lix {\em et al.} \cite{Felix}
and L\"{u}tkenhaus and Jahma \cite{LJ}.

\section{Conclusion}

We have given a simple derivation of beam-splitting and
photon-number splitting attacks on the BB84 protocol in a
realistic implementation: that of using a laser pulse to send
signals instead of the ideal single photon source and of using a
channel that is lossy. From these results we found a bound on the
maximum disturbance or error rate Alice and Bob can accept in
their sifted keys to successfully complete privacy amplification
and give Eve an insignificant amount of knowledge of the key once
this process is complete. These results make no approximation of
small mean photon number.

There is now a large amount of research on eavesdropping on
quantum cryptography. These works have largely neglected analyzing
the EPR based protocols, the first given by Ekert in 1991
\cite{E91}. We think it would be interesting to look at
eavesdropping strategies on realistic implementations of these
protocols.
\begin{acknowledgments}
We would like to thank EPSRC, the European Commission, Elsag SpA
and Hewlett-Packard.
\end{acknowledgments}
\end{document}